\documentclass[twocolumn,showpacs,preprintnumbers,amsmath,amssymb,prb]{revtex4}
\usepackage{graphicx} 
\usepackage{dcolumn} 
\usepackage{bm}
\usepackage{subfigure} 
\usepackage{float} 
\usepackage{color}

\begin{document}

\title{\bf{ Thermodynamics of a one-dimensional frustrated spin-$\frac{1}{2}$ Heisenberg ferromagnet}}

\author{M. H\"{a}rtel}
\author{J. Richter}
\affiliation{Institut f\"{u}r Theoretische Physik, Otto-von-Guericke-Universit\"{a}t Magdeburg, D-39016 Magdeburg, Germany}
\author{D. Ihle}
\affiliation{Institut f\"{u}r Theoretische Physik, Universit\"{a}t Leipzig, D-04109 Leipzig, Germany}
\author{S.-L. Drechsler}
\affiliation{Leibniz-Institut f\"{u}r Festk\"{o}rper- und Werkstoffforschung Dresden, D-01171 Dresden, Germany}

\date{\today}

\begin{abstract}
We calculate the thermodynamic quantities (correlation functions $\langle
{\bf S}_0{\bf S}_n\rangle$,
correlation
length $\xi$,
spin  susceptibility $\chi$, and specific heat $C_V$) of the frustrated one-dimensional 
spin-half $J_1$-$J_2$ Heisenberg ferromagnet, i.e. for $J_2<
0.25|J_1|$, 
using a rotation-invariant
Green's-function 
formalism  and full diagonalization of finite lattices. 
We
find that the critical indices are not changed by $J_2$, i.e., 
$\chi = y_0 T^{-2}$ and $\xi
= x_0T^{-1}$  at $T\to 0$.
However, the
coefficients $y_0$ and $x_0$ linearly
decrease  with increasing $J_2$ according to the 
relations 
$y_0=(1-4J_2/|J_1|)/24$ and 
$x_0 =(1-4J_2/|J_1|)/4$, i.e., both coefficients vanish at $J_2=0.25|J_1|$  
indicating the zero-temperature
phase transition that is accompanied by a change in the low-temperature
behavior of $\chi$ ($\xi$) from $\chi \propto T^{-2}$ ($\xi \propto
T^{-1}$) at $J_2 < 0.25|J_1|$ to $\chi \propto T^{-3/2}$ ($\xi \propto
T^{-1/2}$) at $J_2 = 0.25|J_1|$.
In addition, we detect the existence of an
additional 
low-temperature maximum
in the specific heat when approaching the critical point at $J_2=0.25|J_1|$.
\end{abstract}

\maketitle

\section{ Introduction} \label{intro}
Low-dimensional quantum magnets represent an ideal 
playground to study
systems with strong quantum and thermal fluctuations.\cite{book}  
In particular, much attention has been paid to the one-dimensional (1D) $J_1$-$J_2$ quantum Heisenberg
model, which may serve as a canonical model to study frustration effects
in low-dimensional quantum magnets. Although this model has been studied
frequently (see Ref. \onlinecite{Mikeska04} and references
therein), the model deserves further attention to detect unknown features of
this quantum many-body system, especially in the case of ferromagnetic
nearest-neighbor (NN) interaction $J_1 <
0$.\cite{tonegawa,chubukov,heidrich06,tmrg,krivnov,kuzian,kecke,krivnov08,zinke}
From the experimental side, 
recent studies  have demonstrated that  edge-shared chain
cuprates 
represent a family of quantum magnets for which the 1D $J_1-J_2$ Heisenberg model
is an appropriate starting point for a theoretical description.
Among others, we mention  LiVCuO$_4$, LiCu$_2$O$_2$,
NaCu$_2$O$_2$, Li$_2$ZrCuO$_4$, and Li$_2$CuO$_2$,
\cite{gibson,matsuda,gippius,ender,drechs1,drechs3,drechs4,park,drechsQneu,malek}
which were identified   as quasi-1D frustrated 
spin-$1/2$ magnets 
with a ferromagnetic NN in-chain coupling $J_1<0$ 
and an antiferromagnetic next-nearest-neighbor (NNN)
in-chain coupling
$J_2>0$.
The Hamiltonian of their 1D subsystems considered in this
paper 
is then given by
\begin{equation}  \label{Ham}
  H=J_1\sum_{\langle i,j\rangle}{\bf S}_i{\bf S}_j+J_2\sum_{[ i,j]}{\bf S}_i{\bf S}_j,
\end{equation}
where $\langle i,j\rangle$ runs over the NN and  $[ i,j]$ over the NNN bonds.
For the model (\ref{Ham}) the ferromagnetic  ground state (GS)
gives way for a
singlet GS with spiral correlations at the critical point $J_2 =
0.25|J_1|$.\cite{bader,krivnov}

The edge-shared chain cuprates have attracted much attention due to the 
observation of incommensurate spiral spin ordering at low temperature.
Hence, in these compounds the antiferromagnetic NNN
exchange $J_2$ is strong enough to destroy the ferromagnetic 
GS favored by the
ferromagnetic $J_1$.   
On the other hand, several materials that considered as model systems for 1D
spin-1/2 ferromagnets, 
such as Tetramethylammonium Copper Chloride (
TMCuC [(CH$_3$)$_4$NCuCl$_3$]) (Ref. \onlinecite{TMCuC})
and p-nitrophenyl nitronyl nitroxide (p-NPNN) (C$_{13}$H$_{16}$N$_3$O$_4$),\cite{NPNN} might have also a weak frustrating  
NNN exchange interaction $J_2 <
-0.25J_1$. Moreover, recent investigations suggest that Li$_2$CuO$_2$
is a quasi-1D spin-$1/2$ system with a dominant ferromagnetic $J_1$  and
weak frustrating
antiferromagnetic  $J_2 \approx 0.2
|J_1|$.\cite{malek}

Although for $J_2 <
-0.25J_1$  the GS remains ferromagnetic, 
the frustrating 
$J_2$ may influence the thermodynamics  substantially, in particular near
the zero-temperature critical point at $J_2=0.25|J_1|$.
The investigation of this issue is the aim of this paper. The study 
of the 1D $J_1$-$J_2$ Heisenberg model is faced with the problem that,     
due to the $J_2$ term, neither the Bethe-ansatz solution nor the 
quantum Monte Carlo method is applicable.
Hence we use (i) the full exact diagonalization (ED) of finite
systems of up to $N=22$ lattice sites, 
and (ii) the second-order
Green's-function technique\cite{kondo} that 
has been applied 
recently successfully to
low-dimensional
quantum spin systems.\cite{rgm_new,magfeld,antsyg,ihle_new}
For example,  in Ref.~\onlinecite{magfeld}, by comparison with Bethe-ansatz data
it has been demonstrated
that 
this method leads to qualitatively correct results for the thermodynamics of
the
1D Heisenberg ferromagnet in a magnetic field.  
As the most prominent feature, 
a field-induced 
extra low-temperature maximum in the specific heat has been
found\cite{magfeld} and characterized as a 
peculiar quantum effect.\cite{magfeld,ihle_new}

\section{Full diagonalization of finite lattices} 
Using Schulenburg's 
{\it spinpack} (Ref. \onlinecite{spinpack}) and exploiting
the lattice symmetries and the fact that  $S^z=\sum_i S_i^z$ 
commutes with $H$, 
we are able to calculate 
the exact thermodynamics for periodic chains  of up to $N=22$ spins. 
The comparison of  results for $N=12,14,16,18,20$, and $22$ 
allows to estimate the finite-size effects. The largest matrix
which has to be diagonalized for $N=22$ has $ 29414 \times 29414$ matrix
elements.

\section{Spin-rotation-invariant Green's-function theory} 
To calculate the spin correlation 
functions and the thermodynamic quantities, we determine the transverse 
spin susceptibility 
$\chi_q^{+-}\left(\omega\right)=
-\langle\langle S_q^+;S_{-q}^-\rangle\rangle_\omega$ (here, 
$\langle\langle \ldots;\ldots \rangle\rangle_\omega$ 
denotes the two-time commutator Green's function \cite{elk}) by the spin-rotation-invariant
Green's-function method (RGM). 
\cite{kondo,rgm_new} 
Using the equations of motion up to the second step and supposing rotational symmetry,
i.e., $\langle
S_i^z\rangle=0$, 
we obtain
$
  \omega^2\langle\langle S_q^+;S_{-q}^-\rangle\rangle_\omega=M_q+\langle\langle -\ddot{S}_q^+;S_{-q}^-\rangle\rangle_\omega
$
with $M_q=\langle\left[\left[S_q^+,H\right],S_{-q}^-\right]\rangle$ and $-\ddot{S}_q^+=\left[\left[S_q^+,H\right],H\right]$. 
For the model (\ref{Ham})
the moment $M_q$ is given by the exact expression
\begin{equation}\label{moment}
  M_q=-4\sum_{n=1,2}J_nC_n\left(1-\cos nq\right),
\end{equation}
where $C_n=\langle S_0^+S_n^-\rangle=2\langle S_0^zS_n^z\rangle$. The second derivative $-\ddot{S}_q^+$ 
is approximated as indicated in 
Refs.~\onlinecite{kondo,rgm_new,magfeld,antsyg,ihle_new}. That is, 
in $-\ddot{S}_i^+$ we adopt the decoupling $S_i^+S_j^+S_k^- = 
\alpha\langle S_j^+S_k^-\rangle S_i^++\alpha\langle S_i^+S_k^-\rangle S_j^+$, where in the case 
$J_2<-0.25J_1$ 
with a ferromagnetic GS
the vertex parameter $\alpha$ can be assumed in a good approximation 
to be independent 
of the range of the associated spin
correlators (see the discussion below). We obtain $-\ddot{S}_q^+=\omega_q^2S_q^+$ and
\begin{equation}\label{Greenfunction}
  \chi_q^{+-}\left(\omega\right)= -\langle\langle
S_q^+;S_{-q}^-\rangle\rangle_\omega=\frac{M_q}{\omega_q^2-\omega^2} ,
\end{equation}
with
\begin{equation}\label{dispersion}
\omega_q^2= \hskip-3mm \sum_{n,m(=1,2)} \hskip-3mm
J_nJ_m\left(1-\cos nq\right)
\left[K_{n,m}+4\alpha C_n\left(1-\cos mq\right)\right],
\end{equation}
where $K_{n,n}=1+2\alpha\left(C_{2n}-3C_n\right)$,
$K_{1,2}=2\alpha\left(C_3-C_1\right)$,
and $K_{2,1}=K_{1,2}+4\alpha\left(C_1-C_2\right)$.
\begin{figure}
\includegraphics[height=6cm]{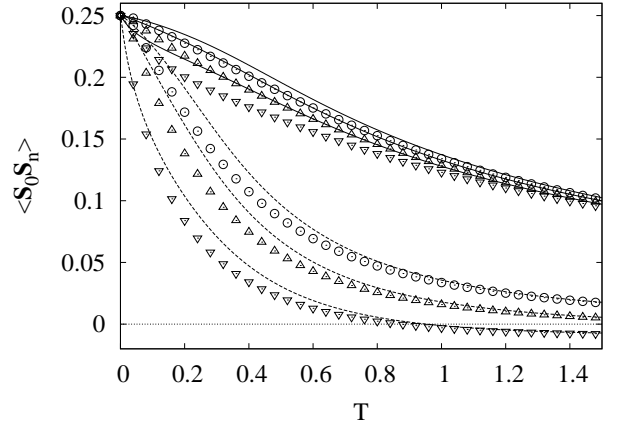}
\caption{\label{korrfuncpic} NN (solid) and NNN (dashed) spin-correlation 
function for $J_2=0,0.1,$ and $0.2$, from top to bottom, calculated by RGM (lines)
and ED (open symbols; N=20).}
\end{figure}
\noindent From the Green's function (\ref{Greenfunction}) the correlation functions $C_n=\frac{1}{N}
\sum_qC_qe^{iqn}$ of arbitrary range $n$ are determined by 
the spectral theorem,\cite{elk}
\begin{equation} \label{spectral}
  C_q=\langle
S_q^+S_{-q}^-\rangle=\frac{M_q}{2\omega_q}\left[1+2n
\left(\omega_q\right)\right],
\end{equation}
where $n(\omega_q)=\left(e^{\omega_q/T}-1\right)^{-1}$ is the Bose function. By the operator identity $S_i^+S_i^-=\frac{1}{2}+S_i^z$ we get the 
sum rule $C_0=\frac{1}{N}\sum_qC_q=\frac{1}{2}$. 
The uniform static 
spin susceptibility $\chi=\lim_{q\to 0}\chi_q$, where 
$\chi_q=\chi_q\left(\omega=0\right)$ and $\chi_q\left(\omega\right)=\frac{1}{2}\chi_q^{+-}\left(\omega\right)$, 
is given by
\begin{equation}\label{susceptibility}
  \chi=-\frac{2}{\Delta}\sum_{n=1,2}n^2J_nC_n\,; \quad \Delta=\hskip-1mm
\sum_{n,m(=1,2)}\hskip-1mm 
n^2 J_nJ_mK_{n,m}.
\end{equation}
The correlation length $\xi$ may be calculated from the expansion 
of the static spin susceptibility around $q=0$ 
(see, e.g., Refs.~\onlinecite{kondo} and \onlinecite{ihle_new}) $\chi_q=\chi/\left(1+\xi^2q^2\right)$.
The ferromagnetic long-range order, occurring in the 1D model at $T=0$ only, 
is described by the condensation 
term $C$ (Ref.~\onlinecite{kondo}) according to $C_n\left(0\right)=\frac{1}{N}\sum_{q(\neq 0)}\left(M_q/2\omega_q\right)e^{iqn}+C$. 
Equating this expression for $n\neq 0$ to the 
exact result 
$C_{n\neq 0}\left(0\right)=\frac{1}{6}$ $\left[\langle\vec S_0\vec S_{n\neq 0}\rangle\left(0\right)=\frac{1}{4}\right]$, 
the ratio $M_q/2\omega_q$ must be 
independent of $q$, because $C_{n\neq 0}$ is independent of $n$. 
This requires the equations $K_{n,m}(0)=0$ [cf. Eqs.~(\ref{moment}) and
(\ref{dispersion})], which yield $\alpha(0)=\frac{3}{2}$. Then, we get $\omega_q(0)=\frac{3}{2}M_q(0)$ and $C=\frac{1}{6}$, where the 
sum rule $C_0=\frac{1}{2}$ is fulfilled. 
In Eq.~(\ref{susceptibility}) we have $\Delta(0)=0$, 
so that $\chi$ diverges as $T\to 0$ 
indicating the ferromagnetic phase transition.

Let us discuss the used assumption that the vertex parameter $\alpha$
is independent of the distance $l$. For that we consider 
an extended decoupling with four different parameters 
$\alpha_l$ ($l=1,\ldots,4$) 
attached to the four correlators $C_l$ appearing in $\omega_q^2$ [cf. Eq.~(\ref{dispersion})]. 
At $T=0$, the four equations $K_{n,m}=0$ ($n,m=1,2$) yield the solutions $\alpha_l(0)=\frac{3}{2}$. 
On the other hand, in the high-temperature limit all vertex parameters approach 
unity.\cite{kondo} Because we have identical vertex parameters at $T=0$ and for $T\to\infty$, we put $\alpha_l=\alpha$ in 
the whole temperature region, as was 
done above.

To evaluate the thermodynamic properties, the correlators $C_l$
($l=1,\ldots,4$) and the vertex parameter $\alpha$ have to be  
determined as numerical solutions of  
a coupled system of five  
non-linear algebraic self-consistency equations for $C_l$ 
including the sum rule $C_0=\frac{1}{2}$ 
according to Eq.~(\ref{spectral}).
Tracing the RGM solution to very low temperature, we find that it
becomes less trustworthy for $J_2$ approaching $J_2 = 0.25|J_1|$. Therefore,
below we will present  
RGM results for $J_2 \le 0.2|J_1|$ only.

\begin{figure}
\includegraphics[height=6cm]{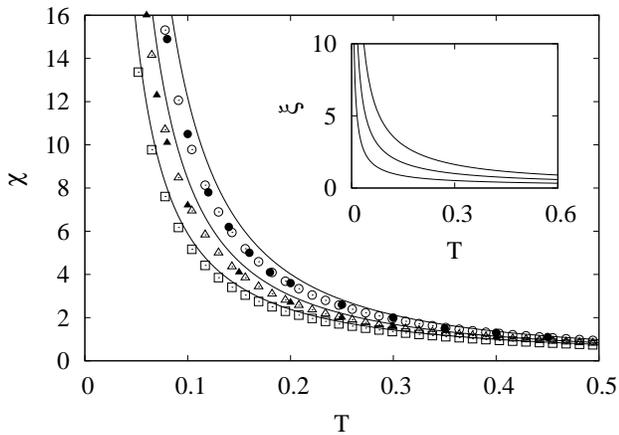}
\caption{\label{susceptchina} Uniform static spin susceptibility calculated 
by RGM (solid lines) and ED (open symbols; N=20) for $J_2=0,0.125,$ and $0.2$, from right to left, and by 
TMRG 
(filled symbols) for $J_2=0$ and $0.125$ (Ref.~\onlinecite{tmrg}). 
The inset shows the correlation length obtained by RGM for $J_2=0,0.125,$ 
and $0.2$, from right to left.}
\end{figure}
\begin{figure}
\includegraphics[height=6cm]{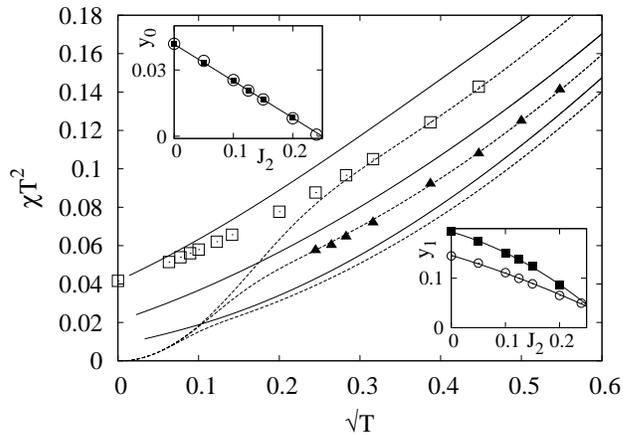}
\caption{$\chi T^2$ versus $\sqrt{T}$ calculated  
by RGM (solid lines) and ED ($N=20$; dashed lines) for 
$J_2=0,0.125,$ and $0.2$, from
top to bottom. For comparison we present also Bethe-ansatz data
(open squares) for $J_2=0$ (Ref.~\onlinecite{yamada1}) and 
TMRG data
(filled triangles) for $J_2=0.125$ (Ref.~\onlinecite{tmrg}). 
The {upper
inset shows the coefficient 
$y_0 = \lim_{T \to 0} \chi T^2$ obtained by}
{the}
{RGM (filled squares) and}
 ED (open circles) in
dependence on $J_2$ as well as a linear fit of the RGM data points (solid
line).
{The lower  inset shows the coefficient $y_1$ [cf.\
Eq.~(\ref{y_0})] obtained by} 
{the}
{RGM (filled squares) and ED (open circles) in
dependence on $J_2$ as well as a quadratic fit of the data points (solid
line).}}
\label{chi_T2}
\end{figure}
\begin{figure}
\includegraphics[height=6cm]{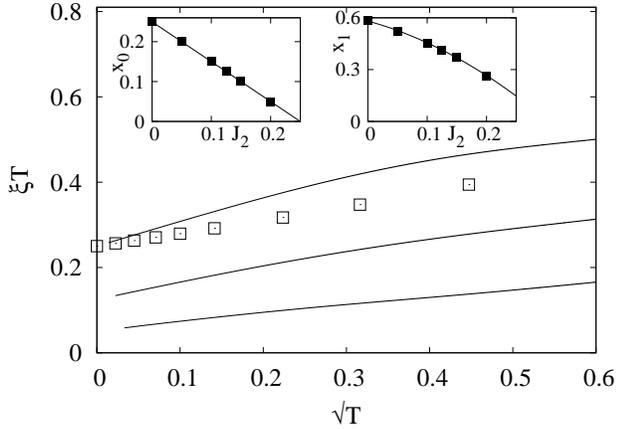}
\caption{ $\xi T$ versus $\sqrt{T}$ by 
{the} RGM (solid lines) for 
$J_2=0,0.125,$ and $0.2$, from top to bottom. 
For comparison we  present also Bethe-ansatz data
(open squares) for $J_2=0$ (Ref.~\onlinecite{yamada2}). 
The {left inset shows the coefficient} 
$x_0 = \lim_{T \to 0} \xi T$ obtained by
{the} 
{RGM (filled squares) in
dependence on $J_2$ as well as a linear fit of the RGM data points (solid
line).}
{The right  inset shows the coefficient 
$x_1$ [cf.\ Eq.~(\ref{x_0})] obtained by}
{the}
{RGM (filled squares) in
dependence on $J_2$ as well as a quadratic fit of the data points (solid
line).}}
\label{xi_T} 
\end{figure}

\section{Results} 

Hereafter, we put $|J_1|=1$. First we consider the NN and 
NNN correlation functions shown in Fig. \ref{korrfuncpic}. 
The RGM results agree qualitatively well with the ED data. Note that the
difference between ED and RGM results at low temperature 
might be partially attributed 
to finite-size effects in the ED data. 
For larger
temperature $T \gtrsim 1$, the agreement becomes perfect.
With increasing 
frustration the correlation functions decrease, where 
the NNN and further-distant 
correlators decay much stronger than the NN correlator
(interestingly, for $J_2=0.2$ 
the NNN correlator changes the sign at $T \approx 1$).   
This frustration effect is reflected in the correlation length $\xi$ depicted in the 
inset of
Fig.~\ref{susceptchina}. At $T=0$, $\xi$ and the uniform static spin susceptibility $\chi$ diverge due to the ferromagnetic GS.
\begin{figure}
\includegraphics[height=6cm]{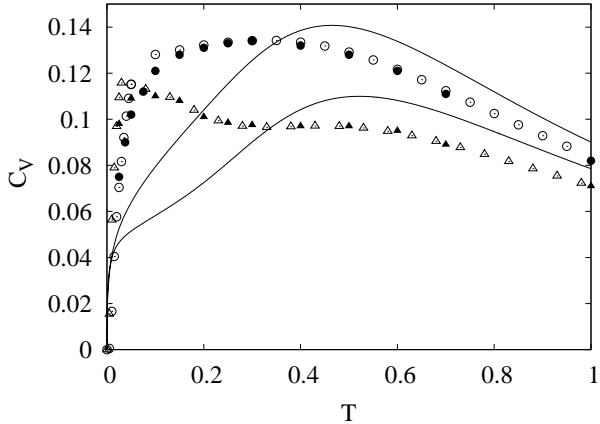}
\caption{Specific heat obtained by RGM (solid lines), ED (open symbols; N=20) and 
TMRG 
(filled symbols; Ref.~\onlinecite{tmrg}) for $J_2=0$ and $0.125$, 
from top to bottom.}\label{specwaermecomp}
\end{figure}
\noindent With growing temperature the decay of $\xi$ increases with 
increasing $J_2$. As shown in Fig.~\ref{susceptchina}, our ED 
data for $\chi$ are in excellent agreement with the results of the 
transfer-matrix renormalization-group (TMRG) study of Ref.~\onlinecite{tmrg} and 
agree well with the RGM results. 
 The susceptibility decreases with increasing $J_2$, 
because this antiferromagnetic interaction counteracts the spin 
orientation along a uniform magnetic field.
\begin{figure}
  \begin{center}
    \includegraphics[height=6cm]{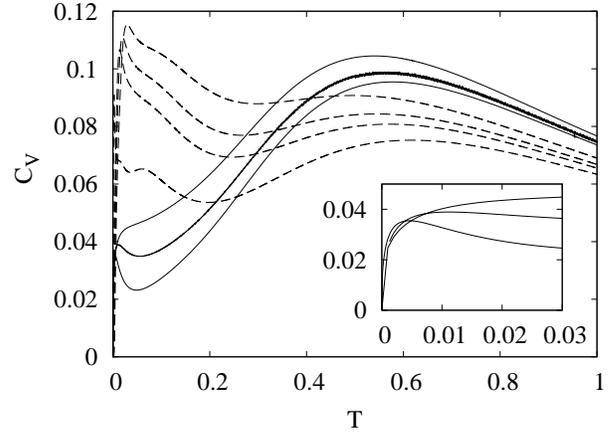}
  \caption{Specific heat calculated by RGM (solid) and ED (dashed curve; N=20) for 
$J_2=0.15,0.18,0.2,$ and $0.24$, from top to bottom. The inset 
exhibits the RGM results in an 
enlarged scale. Note that for $J_2=0.24$ only ED data are shown.
}\label{specheat}
  \end{center}
\end{figure}
\begin{figure}
\includegraphics[height=12cm]{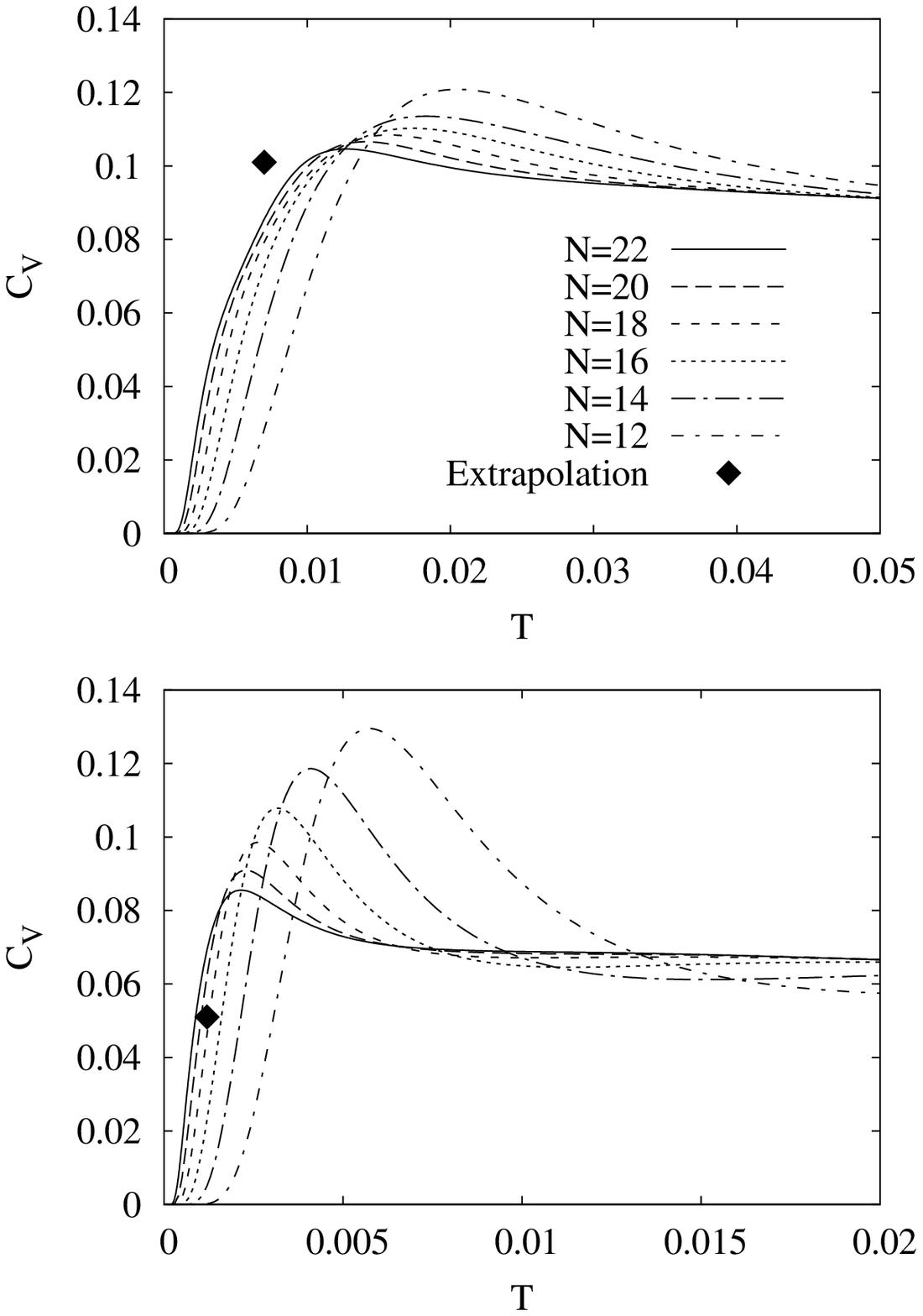}
\caption{\label{c-h0-verschN} Finite-size dependence of the 
low-temperature specific heat for $J_2=0.2$ (upper panel) and $0.24$
(lower panel). 
The lines represent ED data 
for $N=12,14,16,18,20$ and $22$, from top to 
bottom.} 
\end{figure}

Next we investigate the critical behavior of $\chi$ and $\xi$ for $T \to 0$ in more
detail.  
To study the influence of
the frustration on the critical behavior
we follow  
Refs.~\onlinecite{yamada1} and \onlinecite{yamada2}.  The critical indices 
$\gamma$ 
for $\chi$ and
$\nu$ for $\xi$  
can be obtained by analyzing the RGM data for $-\frac{d\log(\chi)}{d\log(T)}$ 
and
$-\frac{d\log(\xi)}{d\log(T)}$ for $T \to 0$.
We find  that $\gamma=2$ and $\nu=1$ are independent of $J_2$ for $J_2 <
0.25$.
Going beyond the leading order in $T$ we know from Bethe-ansatz data\cite{yamada1,yamada2}
and from the renormalization-group technique\cite{kopietz}
that the low-temperature behavior of  
the susceptibility and the correlation length of
the unfrustrated 1D spin-1/2 Heisenberg ferromagnet 
is given by 
\begin{equation}\label{y_0} 
\chi T^2 = y_0 +
y_1 \sqrt{T} + y_2 T+  {\cal O}(T^{3/2}) \;
\end{equation} 
and 
\begin{equation}\label{x_0}
\xi T= x_0 +
x_1 \sqrt{T} + x_2 T +  {\cal O}(T^{3/2})\;.
\end{equation}
Here we adopt this expansion suggested by the existence of the
ferromagnetic
critical
point at $T=0$, but with $J_2$-dependent coefficients for the frustrated
model (\ref{Ham}).
To determine the coefficients $y_0$  and $x_0$, in Figs.~\ref{chi_T2} and \ref{xi_T}  we show
the quantities $\chi T^2$ and $\xi T$ versus $\sqrt{T}$.
Again we find a good agreement of the ED for $\chi T^2$ with Bethe-ansatz
and TMRG data down to quite low temperature. The RGM results for 
$\chi T^2$ and $\xi T$ deviate slightly from the 
Bethe-ansatz
and TMRG data for finite temperature. 
The behavior of the leading coefficients $y_0$ and $x_0$  and the next-order
coefficients $y_1$ and $x_1$   
can be extracted from the data for $\chi T^2$ and $\xi T$ 
by fitting these data 
to Eqs.~(\ref{y_0}) and (\ref{x_0}).
For the RGM we use  
data points up to a cut-off temperature $T=T_{cut}$.
Although we find that the data fit is almost independent of the value of
$T_{cut}$ we choose  $T_{cut}=0.005$, which gives optimal 
coincidence with Bethe-ansatz results available for $J_2=0$ (see below).
On the other hand, the ED data at very low temperature are affected by
finite-size effects. To circumvent this problem 
we proceed as follows. We first determine the temperature $T_{ED}$ down 
to which the
first four digits of the specific heat per site $C_V(T)$ for $N=20$ and $N=22$ coincide.
(We use the specific heat to determine $T_{ED}$, because  
$C_V(T)$ 
is most sensitive to finite-size effects at low temperature,
see also below.) Then we use
the ED data points for  $\chi T^2$
in the temperature region $T_{ED}\le T \le T_{ED} + T_{cut}$ 
to fit them to Eq.~(\ref{y_0}). We find that $T_{ED}$ varies from
$0.22$ at $J_2=0$ to $0.03$ at $J_2=0.24$. 
Obviously, we have to use 
ED data points at higher temperature for the fit 
in comparison to the RGM fit, in particular at small values for $J_2$.
The results for $y_0$ {and $y_1$ as well as for $x_0$ and $x_1$} 
are shown in the insets of Figs.~\ref{chi_T2}
and \ref{xi_T}. It is obvious that the values for $y_0$ determined by RGM and
ED  are very close to each other. 
Note that for the unfrustrated 1D ferromagnet the quantities  $y_0$ and $x_0$
were calculated by the RGM previously in Ref.~\onlinecite{SSI94}.
It was found that $y_0=1/24 \approx 0.041667$ and $x_0=1/4$, which 
agrees with
the Bethe-ansatz results of Refs.~\onlinecite{yamada1} and
\onlinecite{yamada2} [note that $\chi$ defined in Ref.~\onlinecite{yamada1}
is larger by a factor of 4 than $\chi$ given by Eq.~({\ref{susceptibility})].    
Our RGM data confirm these findings (see also Ref.~\onlinecite{ihle_new}).      
The fitting of the ED data at $J_2=0$ yields  $y_0=0.0418$, 
which is still in reasonable agreement
with the Bethe-ansatz result.}    
Including frustration $J_2>0$ we find an almost linear
 decrease in $y_0$ as well in $x_0$ with $J_2$
down to zero at $J_2=0.25$ (cf. the insets of 
Figs.~\ref{chi_T2} and \ref{xi_T}). 
A linear fit of the RGM data points yields 
the relations 
\begin{equation}
 y_0 =(1-4J_2)/24
\quad ; \quad 
x_0 =(1-4J_2)/4 \; ,
\end{equation}
which describe the RGM data in high precision.
The vanishing of $y_0$ and of $x_0$ at $J_2=0.25$ reflects the
zero-temperature phase transition at this point and indicates the change in the
low-temperature behavior of the physical quantities at 
the critical point. 
Using the same $J_2$ data points as in the insets of Figs.~\ref{chi_T2} and
\ref{xi_T}, a polynomial fit according to $y_1 = a_y + b_y J_2+ c_y J_2^2$ ($x_1 =
a_x + b_x J_2+ c_x J_2^2$), indeed,  yields,  {at $J_2=0.25$,} 
finite values 
{$y_1 = 0.047$ for RGM and $y_1 = 0.043$ for ED,  and 
$x_1 = 0.147$} (RGM only).
Hence, our data suggest a change in the low-temperature
behavior of $\chi$ ($\xi$) from $\chi \propto T^{-2}$ ($\xi \propto
T^{-1}$) at $J_2 < 0.25$ to $\chi \propto T^{-3/2}$ ($\xi \propto
T^{-1/2}$) at the zero-temperature critical point
$J_2=0.25$.      
{
Let us mention here again that 
our results for the critical indices $\gamma$ and $\nu$
at $J_2=0.25$ 
are based on the validity of Eqs.~(\ref{y_0}) and
(\ref{x_0}) and the extrapolation of our data from $J_2 < 0.25$ to
$J_2=0.25$. A slightly different index $\gamma$ also being below 
{the  "ferromagnetic" value $\gamma_F=2$ discussed above}, namely
$\gamma=4/3$, is obtained\cite{krivnov_priv}, if
one employs the} {modified spin-wave theory 
by Takahashi \cite{takahashi}}
 {at $J_2=0.25$.}

The next quanti{t}y we consider is the specific heat $C_V$.
In Fig. \ref{specwaermecomp} our RGM and ED results for $C_V$ are compared with 
the 
TMRG data.\cite{tmrg} Obviously, the ED results are in a very good agreement
with the TMRG data. 
The deviation at low temperature, appearing for 
$J_2=0.125$ 
as an increased value of $C_V$ for $0.02 \lesssim T \lesssim 0.1$, 
is ascribed to finite-size effects (see also the discussion below). 
For larger values of $J_2$ the specific heat shows another interesting
low-temperature feature (see Fig. \ref{specheat}). 
In the region $0.125<J_2<0.25$ with a ferromagnetic GS, 
the specific heat exhibits
two maxima. 
Besides the 
broad maximum at $T\approx 0.6$, an additional frustration-induced 
low-temperature maximum 
appears, 
which is found by the ED and RGM methods for $J_2\gtrsim 0.125$ and $\gtrsim 0.16$, respectively. 
As shown by a detailed analysis (see also below), 
the behavior of $C_V$ at very low temperature is appreciably 
affected by finite-size effects. 
In particular, in the ED data, the low-temperature maximum 
is superimposed by a 
quite sharp extra finite-size peak, as can be clearly seen in Fig. 
\ref{specheat} for $J_2=0.24$.
In view of this, the height and the position 
of the true additional low-temperature maximum cannot be extracted unambiguously 
from the ED data, however, its existence is not questioned by this
ambigu{ity}. 
On the other hand, the RGM (see inset of Fig. \ref{specheat}) yields a 
shift of the maximum to lower temperature 
with increasing frustration.

To illustrate the finite-size effects at low temperature, 
in Fig.~\ref{c-h0-verschN} the ED data for the specific heat 
for $J_2=0.2$ and $0.24$
and different chain lengths are plotted. 
As already discussed above,   the first four digits of the $C_V(T)$ data
for $N=20$ and $22$
coincide  down to 
$T_{ED} \approx 0.04$ ($T_{ED}\approx 0.03$) for $J_2=0.2$ ($J_2=0.24$).
(Note
again
that for $J_2=0$ the corresponding value $T_{ED} \approx 0.22$ is much
larger.) Below
$T_{ED}$ finite-size effects become relevant (cf.
Fig.~\ref{c-h0-verschN}).
However, from Fig.~\ref{c-h0-verschN}
it is also evident that the extra low-temperature finite-size peak behaves
monotonously with $N$.
Hence a finite-size extrapolation of the height $c_{peak}$ and the position
$T_{peak}$ of
the extra peak is reasonable. 
We have tested several
extrapolation schemes {and}
found that a three-parameter
fit based on the
formula 
$a(N)=a_0+a_1/N^2 + a_2/N^4$ is well appropriate to extrapolate both
$c_{peak}$ and $T_{peak}$ to $N\to \infty$.
The results of such an extrapolation are shown as filled squares in
Fig.~\ref{c-h0-verschN}. The extrapolated data points indicate that the 
extra peak indeed is a finite-size effect and it vanishes for $N \to \infty$.
However, it is also obvious that the characteristic steep decay of the
specific heat down to $T=0$ starts at lower temperature $T^*$ when approaching the
zero-temperature critical point (we find $T^* \approx 0.05, 0.007$, and $0.002$ for $J_2=0,
0.2$, and $0.24$, respectively).  This behavior is in accordance with the 
shift of
the low-temperature  RGM maximum in $C_V$ mentioned above and is relevant for low-temperature
experiments on quasi-1D ferromagnets.

Finally, let us mention that {in an early paper by Tonegawa and
Harada\cite{tonegawa} and also recently
by Heidrich-Meisner et al.\cite{heidrich06} and Lu et al.\cite{tmrg}} a double-maximum 
structure 
in $C_V$ was already found for $0.25 \le J_2 \lesssim 0.4$, however, with a
low-temperature maximum that becomes much more pronounced approaching the
critical point.
In this case,  
the low-temperature peak in $C_V(T)$ was ascribed to excitations from a singlet GS to a low-lying 
ferromagnetic multiplet.\cite{heidrich06} In our case $J_2<0.25$. Above the
fully polarized ferromagnetic GS multiplet many low-lying multiplets exist, and
the appearance of the additional low-temperature maximum is
attributed to a more subtle interplay 
between all of these low-lying states.

\section{Summary} 
In this paper we explored the influence of the NNN coupling 
$J_2 \le 0.25|J_1|$ on the thermodynamic properties of 
the 1D spin-1/2 Heisenberg ferromagnet using ED and RGM methods. 
The results of both methods are in qualitatively good agreement. 
We found that the critical behavior of the susceptibility $\chi$ and
the correlation length $\xi$ is not changed by the frustrating
$J_2$. However, $\lim_{T \to 0} \chi T^2$ and $\lim_{T \to 0} \xi T$ go to
zero for $J_2 \to 0.25|J_1|$ indicating a change in the low-temperature
behavior of $\chi$ ($\xi$) from $\chi \propto T^{-2}$ ($\xi \propto
T^{-1}$) at $J_2 < 0.25|J_1|$ to $\chi \propto T^{-3/2}$ ($\xi \propto
T^{-1/2}$) at the 
critical point $J_2 = 0.25|J_1|$.
Another interesting feature is the appearance of a double-maximum structure 
in the 
specific heat $C_V$, where the additional frustration-induced low-temperature maximum was found by ED (RGM) to occur for 
$J_2/|J_1|\gtrsim 0.125$ $(0.16)$.\\

{{\it   Acknowledgment:}
This work was supported by the DFG (projects No. RI615/16-1 and No. DR269/3-1).}
One of us (S.-L. D.) is indebted to V.Ya.~Krivnov for useful discussions.
{Further discussions with S.~Sachdev and A.~Zvyagin are kindly
acknowledged.}

\end{document}